\documentclass[11pt,a4paper]{article}%
\usepackage{amsfonts}
\usepackage{amsmath}
\usepackage{amssymb}
\usepackage{graphicx}
\usepackage{geometry}%
\providecommand{\U}[1]{\protect\rule{.1in}{.1in}}
\newtheorem{theorem}{Theorem}

\newtheorem{corollary}{Corollary}

\newtheorem{definition}{Definition}

\newtheorem{proposition}{Proposition}

\makeatletter
\def \@removefromreset#1#2{\let \@tempb \@elt
\def \@tempa#1{@&#1}\expandafter \let \csname @*#1*\endcsname \@tempa
\def \@elt##1{\expandafter \ifx \csname @*##1*\endcsname \@tempa \else
\noexpand \@elt{##1}\fi}     \expandafter \edef \csname cl@#2\endcsname{\csname cl@#2\endcsname}     \let \@elt \@tempb
\expandafter \let \csname @*#1*\endcsname \@undefined}

\@removefromreset{equation}{section}

\@removefromreset{theorem}{section}
\makeatother
\begin{document}

\title{Violation of general Bell inequalities by a pure bipartite quantum state}
\author{Elena R. Loubenets$^{1,2}$ and Min Namkung$^{1,3}$\\$^{1}$National Research University Higher School of Economics, \\Moscow 101000, Russia \\$^{2}$ Steklov Mathematical Institute of Russian Academy of Sciences,\\ Moscow 119991, Russia\\ $^{3}$Department of Applied Mathematics and Institute of Natural Sciences,\\Kyung Hee University, Yongin 17104, Republic of Korea}
\maketitle

\begin{abstract}
In the present article, based on the formalism introduced in [Loubenets,
\emph{J. Math. Phys}. \textbf{53}, 022201 (2012)], we derive for a pure
bipartite quantum state a new upper bound on its maximal violation of general
Bell inequalities. This new bound indicates that, for an infinite dimensional
pure bipartite state with a finite sum of its Schmidt coefficients, violation
of any general Bell inequality is bounded from above by the value independent
on a number of settings and a type of outcomes, continuous or discrete,
specific to this Bell inequality. As an example, we apply our new general
results to specifying upper bounds on the maximal violation of general Bell
inequalities by infinite dimensional bipartite states having the Bell states
like forms comprised of two binary coherent states $|\alpha\rangle
,|-\alpha\rangle$, with $\alpha>0$. We show that, for each of these bipartite
coherent states, the maximal violation of general Bell inequalities cannot
exceed the value $3$ and analyse numerically the dependence of the derived
analytical upper bounds on a parameter $\alpha>0$.

\end{abstract}

\section{Introduction}

Ever since the seminal paper of Bell \cite{bell} quantum violation of Bell
inequalities was analysed, analytically and numerically, in many papers and is
now used in many quantum information processing tasks. The maximal quantum
violation of the Clauser-Horne-Shimony-Holt (CHSH) inequality
\cite{j.f.clauser} is well known
\cite{b.s.cirelson,b.s.tsirelson,r.f.werner,scarani} and constitutes $\sqrt
{2}$, for any bipartite quantum state, possibly infinite dimensional. It was
also recently proved \cite{original1, original2} that the maximal quantum
violation of the original Bell inequality\footnote{For the original Bell
inequality see also \cite{loubenets11}.} \cite{bell} is equal to $\frac{3}%
{2}.$

More generally, quantum violation of any (unconditional) \emph{correlation}
bipartite Bell inequality cannot\footnote{This follows from the definition of
the Grothendieck's constant $K_{G}^{(\mathbb{R})}$\ and Theorem 2.1 in
\cite{b.s.tsirelson}.} exceed the real Grothendiek's constant $K_{G}%
^{(\mathbb{R})}\in\lbrack1.676,1.783]$ but this is not already the case for
quantum violation of general \cite{e.r.loubenets1} bipartite Bell
inequalities, in particular, bipartite Bell inequalities on joint
probabilities, and last years this problem was intensively discussed in the
literature\ within different mathematical approaches
\cite{7,8,e.r.loubenets2,10,11,12,14,15,e.r.loubenets4,e.r.loubenets5,e.r.loubenets6}%
.

For an arbitrary bipartite state $\rho$ on a Hilbert space $\mathcal{H}%
_{1}\otimes\mathcal{H}_{2},$ with dimensions $d_{n}:=\dim\mathcal{H}_{n},$ the
upper bound on the maximal violation of general bipartite Bell inequalities
for $S_{n}$ measurements with any type of outcomes at each $n=1,2$ site
follows from the general upper bounds derived in
\cite{e.r.loubenets2,e.r.loubenets4,e.r.loubenets5,e.r.loubenets6} for an
$N$-partite quantum state and reads\footnote{For bound (\ref{0_1}), see Eq.
(64) in \cite{e.r.loubenets2}.}
\begin{equation}
2\min\{d_{1},d_{2},S_{1},S_{2}\}-1\label{0_1}%
\end{equation}
-- in case of generalized quantum measurements and\footnote{For bound
(\ref{0_2}), see Eq. (22) in \cite{e.r.loubenets4} and Eq. (31) in
\cite{e.r.loubenets6}.}%
\begin{align}
\min\left\{  \sqrt{d},3\right\}  ,\ \ \ \text{for}\ S &  =2,\label{0_2}\\
\min\left\{  \sqrt{d^{S}},\text{ }2\min\{d,S\}-1\right\}  ,\ \ \ \text{for}\ S
&  \geq3,\nonumber
\end{align}
-- in the case $d_{1}=d_{2}=d,$ $S_{1}=S_{2}=S$ and projective quantum
measurements at both sites.

From bound (\ref{0_1}) it follows that, for an arbitrary two-qudit state, pure
or mixed, violation of any general Bell inequality cannot exceed the value%
\begin{equation}
2\min\{d_{1},d_{2}\}-1, \label{0_3}%
\end{equation}
while, for an arbitrary infinite dimensional bipartite quantum state, pure or
mixed, violation of any general Bell inequality with $S_{1},S_{2}\geq1$
settings at the corresponding sites cannot exceed
\begin{equation}
2\min\{S_{1},S_{2}\}-1. \label{0_4}%
\end{equation}

For the two-qudit state$\frac{1}{\sqrt{d}}\sum_{j=1}^{d}|j\rangle^{\otimes2}$,
it was also found in \cite{11} (see there Theorem 0.3) via the operator space
formalism that the maximal quantum violation of all general Bell inequalities
cannot exceed $C\frac{d}{\sqrt{\ln d}},$ where $C$ is an unknown constant
independent on a dimension $d$.

In the present article, based on the local quasi hidden variable (LqHV)
formalism introduced and developed in
\cite{e.r.loubenets2,e.r.loubenets4,e.r.loubenets5,e.r.loubenets6,e.r.loubenets2-1,e.r.loubenets3}%
, we derive a new upper bound on the maximal violation of general Bell
inequalities by\emph{ an arbitrary pure bipartite quantum state}, possibly
infinite dimensional.

This new upper bound, expressed in terms of the Schmidt coefficients for a
pure bipartite state, is consistent with the general upper bound \ref{0_1}
valid for an arbitrary bipartite state, pure or mixed, and indicates that, for
an infinite dimensional pure bipartite state with a finite sum of its Schmidt
coefficients, violation of any general Bell inequality is bounded from above
by the value independent on a number of settings and a type of outcomes,
continuous or discrete, specific to this Bell inequality.

As an example, we apply our new result to finding an upper bound on the
maximal violation of general Bell inequalities by infinite dimensional
entangled bipartite states having the Bell states like forms comprised of two
binary coherent states $|\alpha\rangle,$ $|-\alpha\rangle,$ $\alpha>0.$ These
entangled bipartite coherent states have been intensively discussed in the
literature
\cite{x.wang,h.jeong,h.kwon,h.jeong2,h.kwon2,c.-y.park,a.serafini,m.namkung}
in view of their experimental implementations. We show that, for each of these
bipartite coherent states, the maximal violation of all general Bell
inequalities cannot exceed the value $3$ and analyse numerically the
dependence of the derived upper bound on a parameter $\alpha>0.$

The article is organized as follows.

In Section 2, we recall (in relation to a bipartite case) the notion of a
general Bell inequality \cite{e.r.loubenets1} and describe the state
parameters \cite{e.r.loubenets2} characterizing (Theorem 1) the maximal
violation of general Bell inequalities by an arbitrary quantum state.

In Section 3, we specify the notion of a source operator for a bipartite state
(introduced for a general $N$-partite state in
\cite{e.r.loubenets2,e.r.loubenets8}) and present (Theorem 2) the analytical
upper bound on the maximal violation of general Bell inequalities by an
arbitrary bipartite state, which is expressed in terms of source operators of
this state and follows from the more tight upper bound, presented by Eq. (53)
in \cite{e.r.loubenets2}.

In Section 4, for the maximal violation of general Bell inequalities by a
\emph{pure} bipartite state, we derive \emph{a new upper bound} (Theorem 3)
expressed in terms of the Schmidt coefficients of this pure state, and discuss
the main consequences (Corollaries 1,2) following from this new result.

In Section 5, we apply our new results to specifying upper bounds (Proposition
1) on the maximal violations of general Bell inequalities by infinite
dimensional pure bipartite states, having the Bell states like forms comprised
of two binary coherent states $|\alpha\rangle,$ $|-\alpha\rangle$, $\alpha>0.$
We prove that, for each of these bipartite coherent states, the maximal
violation of general Bell inequalities cannot exceed the value $3$ and analyse
numerically the dependence of the derived upper bounds on a parameter
$\alpha>0.$

In Section 6, we summarize the main results of the present article.

\section{Preliminaries: quantum violation of Bell inequalities}

Consider\footnote{On the probabilistic description of a general correlation
scenario, see \cite{e.r.loubenets0}.} a general bipartite correlation scenario
where each of two participants performs $S_{n}\geq1$, $n=1,2,$ different
measurements, indexed by numbers $s_{n}=1,...,S_{n}$ and with outcomes
$\lambda_{n}\in\Lambda_{n}$. We refer to this correlation scenario as
$S_{1}\times S_{2}$-setting and denote by $\mathrm{P}_{(s_{1},s_{{2}})}%
(\cdot)$ the probability distribution of outcomes $(\lambda_{1},\lambda
_{2})\in\Lambda:=\Lambda_{1}\times\Lambda_{2}$ under the joint measurement
specified by a tuple $(s_{1},s_{2})$ of settings where each $n$-th participant
performs a measurement $s_{n}$ at the $n$-th site. The complete probabilistic
description of such an $S_{1}\times S_{2}$-setting correlation scenario is
given by the family
\begin{align}
\mathcal{P}_{S,\Lambda}  &  :=\left\{  \mathrm{P}_{(s_{1},s_{{2}})}\mid
s_{n}=1,...,S_{n},\text{\ \ \ }n=1,2\right\}  ,\label{1}\\
S  &  :=S_{1}\times S_{2}.\nonumber
\end{align}
of joint probability distributions.

A correlation scenario admits a local hidden variable (LHV) model\footnote{For
the definition of this notion under a general correlation scenario, see
Definition 4 \cite{e.r.loubenets0}.} if each of its joint probability
distributions $\mathrm{P}_{(s_{1},s_{2})}\in\mathcal{P}_{S,\Lambda}$ admits
the representation%
\begin{equation}
\mathrm{P}_{(s_{1},s_{2})}\left(  \mathrm{d}\lambda_{1}\times\mathrm{d}%
\lambda_{2}\right)  =%
{\displaystyle\int\limits_{\Omega}}
P_{1,s_{1}}(\mathrm{d}\lambda_{1}|\omega)\cdot P_{2,s_{{2}}}(\mathrm{d}%
\lambda_{2}|\omega)\nu(\mathrm{d}\omega) \label{2}%
\end{equation}
in terms of a unique probability distribution $\nu(\mathrm{d}\omega)$ of some
variables $\omega\in\Omega$ and conditional probability distributions
$P_{n,s_{n}}(\mathrm{\cdot}|\omega),$ referred to as ``local'' in the sense
that each $P_{n,s_{n}}(\mathrm{\cdot}|\omega)$ at $n$-th site depends only on
a measurement $s_{n}=1,...,S_{n}$ at an $n$-th site.

Under an $S_{1}\times S_{2}$-setting correlation scenario described by a
family of joint probability distributions (\ref{1}), consider a linear
combination \cite{e.r.loubenets1}
\begin{equation}
\mathcal{B}_{\Phi_{S,\text{ }\Lambda}}(\mathcal{P}_{S,\text{ }\Lambda}%
):=\sum_{s_{1},s_{{2}}}\left\langle \text{ }\phi_{(s_{1},s_{2})}(\lambda
_{1},\lambda_{2})\right\rangle _{\mathrm{P}_{(s_{1},s_{2})}} \label{3}%
\end{equation}
of the mathematical expectations
\begin{equation}
\left\langle \text{ }\phi_{(s_{1},s_{2})}(\lambda_{1},\lambda_{2}%
)\right\rangle _{\mathrm{P}_{(s_{1},s_{2})}}:=\int\limits_{\Lambda}%
\phi_{(s_{1},s_{2})}(\lambda_{1},\lambda_{2})\text{ }\mathrm{P}_{(s_{1}%
,s_{2})}(\mathrm{d}\lambda_{1}\times\mathrm{d}\lambda_{2}) \label{4}%
\end{equation}
of an arbitrary form, specified by a collection%
\begin{equation}
\Phi_{S,\Lambda}=\left\{  \phi_{(s_{{1}},s_{{2}})}:\Lambda\rightarrow
\mathbb{R}\mid s_{n}=1,...,S_{n};\text{ \ \ }n=1,2\right\}  \label{5}%
\end{equation}
of bounded real-valued functions $\phi_{(s_{{1}},s_{{2}})}$ on $\Lambda
=\Lambda_{1}\times\Lambda_{2}.$ Depending on a choice of a bounded function
$\phi_{(s_{{1}},s_{{2}})}$ and types of outcome sets $\Lambda_{n}$, $n=1,2,$
expression (\ref{4}) can constitute either the probability of some observed
event or if $\Lambda_{n}\subset\mathbb{R},$ $n=1,2,$ the mathematical
expectation (mean) of the product of observed outcomes (called in quantum
information as a correlation function) or have a more complicated form.

If an $S_{1}\times S_{2}$-setting correlation scenario (\ref{1}) admits the
LHV modelling in the sense of representation (\ref{2}), then every linear
combination (\ref{3}) of its mathematical expectations (\ref{4}) satisfies the
\textquotedblleft tight\textquotedblright\ LHV constraints
\cite{e.r.loubenets1,e.r.loubenets2}
\begin{align}
\mathcal{B}_{\Phi_{S,\Lambda}}^{\inf}  &  \leq\mathcal{B}_{\Phi_{S,\Lambda}%
}(\mathcal{P}_{S,\Lambda})|_{lhv}\leq\mathcal{B}_{\Phi_{S,\Lambda}}^{\sup
},\label{6}\\[0.03in]
\left\vert \text{ }\mathcal{B}_{\Phi_{S,\Lambda}}(\mathcal{P}_{S,\Lambda
})\right\vert _{lhv}  &  \leq\mathcal{B}_{\Phi_{S,\Lambda}}^{lhv}%
:=\max\left\{  \left\vert \mathcal{B}_{\Phi_{S,\Lambda}}^{\sup}\right\vert
,\left\vert \mathcal{B}_{\Phi_{S,\Lambda}}^{\inf}\right\vert \right\}
,\nonumber
\end{align}
where constants
\begin{align}
\mathcal{B}_{\Phi_{S,\Lambda}}^{\sup}  &  :=\sup_{\mathcal{P}_{S,\Lambda}%
\in\mathfrak{G}_{S,\Lambda}^{lhv}}\mathcal{B}_{\Phi_{S,\Lambda}}%
(\mathcal{P}_{S,\Lambda})\label{7}\\[0.03in]
&  =\sup_{\lambda_{n}^{(s_{n})}\in\Lambda_{n},\text{ }\forall s_{n},\text{
}n=1,2}\ \sum_{s_{1},s_{{2}}}\phi_{(s_{1},s_{2})}(\lambda_{1}^{(s_{1}%
)},\lambda_{2}^{(s_{2})}),\nonumber\\
\mathcal{B}_{\Phi_{S,\Lambda}}^{\inf}  &  :=\inf_{\mathcal{P}_{S,\Lambda}%
\in\mathfrak{G}_{S,\Lambda}^{lhv}}\mathcal{B}_{\Phi_{S,\Lambda}}%
(\mathcal{P}_{S,\Lambda})\nonumber\\
&  =\inf_{\lambda_{n}^{(s_{n})}\in\Lambda_{n},\text{ }\forall s_{n},\text{
}n=1,2}\ \sum_{s_{1},s_{{2}}}\phi_{(s_{1},s_{2})}(\lambda_{1}^{(s_{1}%
)},\lambda_{2}^{(s_{2})}).\nonumber
\end{align}
\medskip Here, $\mathfrak{G}_{S,\Lambda}^{lhv}$ denotes the set of all
families (\ref{1}) of joint probability distributions describing $S_{1}\times
S_{2}$-setting correlation scenarios with outcomes in $\Lambda=\Lambda
_{1}\times\Lambda_{2}$ admitting the LHV modelling.

Depending on a form of functional (\ref{3}), which is specified by a family
$\Phi_{S,\Lambda}$ of bounded functions (\ref{6}), some of the LHV constraints
in (\ref{6}) can hold for a wider (than LHV) class of correlation scenarios,
some may be simply trivial, i.e. fulfilled under all correlation scenarios.

\begin{definition}
\cite{e.r.loubenets1,e.r.loubenets2}\emph{\ }Each of the tight linear
LHV\ constraints in (\ref{6}) that can be violated under a non-LHV correlation
scenario is referred to as a general Bell inequality.
\end{definition}

Bell inequalities on correlation functions (like the CHSH inequality) and Bell
inequalities on joint probabilities constitute particular classes of general
Bell inequalities. The general form of $N$-partite \emph{Bell inequalities} on
correlation functions (called \emph{correlation Bell inequalities)} and the
general form of $N$-partite Bell inequalities on joint probabilities are
introduced in \cite{e.r.loubenets1}.

If, under a bipartite correlation scenario, all joint measurements
$(s_{1},s_{{2}})$ are performed on a quantum state $\rho$ on a Hilbert space
$\mathcal{H}_{1}\otimes\mathcal{H}_{2}$, then each joint probability
distribution $\mathrm{P}_{(s_{1},s_{{2}})}$ in (\ref{1})\ takes the form
\begin{equation}
\mathrm{P}_{(s_{1},s_{{2}})}(\mathrm{d}\lambda_{1}\times\mathrm{d}\lambda
_{2})=\mathrm{tr}[\rho\{\mathrm{M}_{1}^{(s_{1})}(\mathrm{d}\lambda_{1}%
)\otimes\mathrm{M}_{2}^{(s_{{2}})}(\mathrm{d}\lambda_{2})\}],\label{8}%
\end{equation}
where $\mathrm{M}_{n}^{(s_{n})}(\cdot),$ $\mathrm{M}_{n}^{(s_{n})}(\Lambda
_{n})=\mathbb{I}_{\mathcal{H}_{n}},$ is a normalized positive operator-valued
(POV) measure, describing $s_{n}$-th quantum measurement at $n$-th site. For
this correlation scenario, we denote the family (\ref{1}) of joint probability
distributions by%
\begin{equation}
\mathcal{P}_{S,\Lambda}^{(\rho,\mathfrak{m}_{S,\Lambda})}:=\left\{
\mathrm{tr}[\rho\{\mathrm{M}_{1}^{(s_{1})}(\mathrm{d}\lambda_{1}%
)\otimes\mathrm{M}_{2}^{(s_{{2}})}(\mathrm{d}\lambda_{2})\}],\text{ }%
s_{n}=1,...,S_{n},\text{ }n\in1,2\right\}  ,\label{9}%
\end{equation}
where
\begin{equation}
\mathfrak{m}_{S,\Lambda}:=\left\{  \mathrm{M}_{n}^{(s_{n})}\mid\text{ }%
s_{n}=1,...,S_{n},\text{ }n\in1,2\right\}  \label{10}%
\end{equation}
is the collection of all local POV measures at two sites, describing this
quantum correlation scenario.

The following statement follows if $N=2$ from the general statements for an
$N$-partite case in \cite{e.r.loubenets2} (see there Eq. (48) and Lemma 3).

\begin{theorem}
\cite{e.r.loubenets2} For an $S_{1}\times S_{2}$-setting quantum correlation
scenario\ (\ref{9}) performed\emph{ }on a state $\rho$ on $\mathcal{H}%
_{1}\otimes\mathcal{H}_{2}$, possibly infinite dimensional, every linear
combination (\ref{3}) of its mathematical expectations (\ref{4}) satisfies the
\textquotedblleft tight\textquotedblright\ constraints:%
\begin{align}
&  \mathcal{B}_{\Phi_{S,\Lambda}}^{\inf}-\frac{\mathrm{\Upsilon}_{S_{1}\times
S_{2}}^{(\rho,\Lambda)}-1}{2}(\mathcal{B}_{\Phi_{S,\Lambda}}^{\sup
}-\mathcal{B}_{\Phi_{S,\Lambda}}^{\inf})\nonumber\\
&  \leq\mathcal{B}_{\Phi_{S,\Lambda}}(\mathcal{P}_{S,\Lambda}^{\rho
,\mathfrak{m}_{S,\Lambda}})\label{11}\\
&  \leq\mathcal{B}_{\Phi_{S,\Lambda}}^{\sup}+\frac{\mathrm{\Upsilon}%
_{S_{1}\times S_{2}}^{(\rho,\Lambda)}-1}{2}(\mathcal{B}_{\Phi_{S,\Lambda}%
}^{\sup}-\mathcal{B}_{\Phi_{S,\Lambda}}^{\inf}),\nonumber
\end{align}
where
\begin{equation}
1\leq\mathrm{\Upsilon}_{S_{1}\times S_{2}}^{(\rho,\Lambda)}:=\sup_{_{\text{
}\mathfrak{m}_{S,\Lambda},\Phi_{S,\Lambda},\mathcal{B}\text{ }_{\Phi
_{S,\Lambda}}^{lhv}\neq0}}\frac{\left\vert \mathcal{B}_{\Phi_{S,\Lambda}%
}(\mathcal{P}_{S,\Lambda}^{\rho,\mathfrak{m}_{S,\Lambda}})\right\vert
}{\mathcal{B}_{\Phi_{S,\Lambda}}^{lhv}} \label{12}%
\end{equation}
is the maximal violation by a state $\rho$\ of all $S_{1}\times S_{2}$-setting
general Bell inequalities with outcomes $(\lambda_{1},\lambda_{2})\in\Lambda$.
\end{theorem}

\section{General upper bounds}

Let $T_{S_{1}\times S_{2}}^{(\rho)}$ be a self-adjoint trace class dilation of
a state $\rho$ on $\mathcal{H}_{1}\otimes\mathcal{H}_{2}$ to the Hilbert space
$\mathcal{H}_{1}^{\otimes S_{1}}\otimes\mathcal{H}_{2}^{\otimes S_{2}}$. By
its definition%
\begin{align}
&  \mathrm{tr}\left[  T_{S_{1}\times S_{2}}^{(\rho)}\left\{  \mathbb{I}%
_{\mathcal{H}_{1}^{\otimes k_{1}}}\otimes X_{1}\otimes\mathbb{I}%
_{\mathcal{H}_{1}^{\otimes(S_{1}-1-k_{1})}}\otimes\mathbb{I}_{\mathcal{H}%
_{2}^{\otimes k_{2}}}\otimes X_{2}\otimes\mathbb{I}_{\mathcal{H}_{2}%
^{\otimes(S_{2}-1-k_{2})}}\right\}  \right] \label{13}\\
&  =\mathrm{tr}\left[  \rho\left\{  X_{1}\otimes X_{2}\right\}  \right]
,\text{ \ \ }k_{n}=0,...,(S_{n}-1),\text{ \ \ }n=1,2,\nonumber
\end{align}
for all bounded operators $X_{n}$ on $\mathcal{H}_{n}$, $n=1,2.$ Clearly,
$T_{1\times1}^{(\rho)}=\rho$, $\mathrm{tr}[T_{S_{1}\times S_{2}}^{(\rho
)}]=1\ $and $\left\Vert T_{S_{1}\times S_{2}}^{(\rho)}\right\Vert _{1}\geq1,$
where $\left\Vert \cdot\right\Vert _{1}$ means the trace norm. In
\cite{e.r.loubenets2,e.r.loubenets8}, we call a self-adjoint trace class
operator $T_{S_{1}\times S_{2}}^{(\rho)}$ as \emph{an}\textrm{ }$S_{1}\times
S_{2}$-\emph{setting source operator for a state} $\rho$ on $\mathcal{H}%
_{1}\otimes\mathcal{H}_{2}.$ As proved\footnote{See Proposition 1 in
\cite{e.r.loubenets2} for a general $N$-partite case.} in
\cite{e.r.loubenets2}, for every state $\rho$ on $\mathcal{H}_{1}%
\otimes\mathcal{H}_{2}$ and arbitrary integers\emph{\ }$S_{1},S_{2}\geq1$, a
source operator $T_{S_{1}\times S_{2}}^{(\rho)}$ exists.

For a separable quantum state, there always exists a positive source operator.
However, for an arbitrary bipartite quantum state, a source operator
$T_{S_{1}\times S_{2}}^{(\rho)}$ does not need to be either positive or, more
generally, tensor positive. The latter general notion introduced in
\cite{e.r.loubenets2} means that
\begin{equation}
\mathrm{tr}\left[  T_{S_{1}\times S_{2}}^{(\rho)}\left\{  \text{ }A_{1}%
\otimes\cdots\otimes A_{S_{1}}\otimes B_{1}\otimes\cdots\otimes B_{S_{2}%
}\right\}  \right]  \geq0
\end{equation}
for all positive bounded operators $A_{k},B_{m}$ on $\mathcal{H}_{1}$ and
$\mathcal{H}_{2},$ respectively.

Theorem 3 in \cite{e.r.loubenets2} implies.

\begin{theorem}
For an arbitrary state $\rho$ on a complex Hilbert space $\mathcal{H}%
_{1}\otimes\mathcal{H}_{2}$, possibly infinite dimensional, and any integers
$S_{n}\geq1,$ $n=1,2,$ the maximal violation
\begin{equation}
\mathrm{\Upsilon}_{S_{1}\times S_{2}}^{(\rho)}:=\sup_{\Lambda}\mathrm{\Upsilon
}_{S_{1}\times S_{2}}^{(\rho,\Lambda)} \label{14}%
\end{equation}
by a state $\rho$\ of all $S_{1}\times S_{2}$-setting general Bell
inequalities for any type of outcomes, discrete or continuous, at each of two
sites satisfies the relation%
\begin{align}
1  &  \leq\mathrm{\Upsilon}_{S_{1}\times S_{2}}^{(\rho)}\leq\min\left\{
\inf_{T_{S_{1}\times1}^{(\rho)}}\left\Vert T_{S_{1}\times1}^{(\rho
)}\right\Vert _{1},\inf_{T_{1\times S_{2}}^{(\rho)}}\left\Vert T_{1\times
S_{2}}^{(\rho)}\right\Vert _{1}\right\} \label{15}\\
&  \leq\inf_{T_{S_{1}\times S_{2}}^{(\rho)}}\left\Vert T_{S_{1}\times S_{2}%
}^{(\rho)}\right\Vert _{1},\nonumber
\end{align}
where $T_{S_{1}\times1}^{(\rho)},$ $T_{1\times S_{2}}^{(\rho)}$ and
$T_{S_{1}\times S_{2}}^{(\rho)}$ are source operators of a state $\rho$ on
Hilbert spaces $\mathcal{H}_{1}^{\otimes S_{1}}\otimes\mathcal{H}_{2},$
$\mathcal{H}_{1}\otimes\mathcal{H}_{2}^{\otimes S_{2}}$ and $\mathcal{H}%
_{1}^{\otimes S_{1}}\otimes\mathcal{H}_{2}^{\otimes S_{2}},$ respectively.
\end{theorem}

If for a state $\rho,$ the maximal violation parameter $\mathrm{\Upsilon
}_{S_{1}\times S_{2}}^{(\rho)}$ is bounded from above by a value independent
on numbers $S_{1} ,S_{2} \geq1$ of settings at each of two sites, then the
parameter \cite{e.r.loubenets4,e.r.loubenets5}%
\begin{equation}
1\leq\mathrm{\Upsilon}_{\rho}:=\sup_{S_{1},S_{2}}\mathrm{\Upsilon}%
_{S_{1}\times S_{2}}^{(\rho)} \label{16_}%
\end{equation}
constitutes the maximal violation by a state $\rho$\ of \emph{all} possible
general Bell inequalities.

The general upper bounds on parameter $\mathrm{\Upsilon}_{S_{1}\times S_{2}%
}^{(\rho)}$ valid for an arbitrary state $\rho,$ pure or mixed, are presented
by relations (\ref{0_1}) in Introduction. In the following Section, based on
Theorem 2, we derive for the maximal quantum violation of all general
$S_{1}\times S_{2}$-setting Bell inequalities, a new upper bound which is
consistent with bound (\ref{0_1}) but is true only for pure bipartite states.

\section{Upper bounds for a pure bipartite state}

Recall that, for any pure bipartite state $|\psi\rangle\langle\psi|$ on
$\mathcal{H}_{1}\otimes\mathcal{H}_{2}$, the non-zero eigenvalues
$0<\lambda_{k}(\psi)\leq1$ of its reduced states on $\mathcal{H}_{1}$ and
$\mathcal{H}_{2}$ coincide and have the same multiplicity while vector
$|\psi\rangle\in\mathcal{H}_{1}\otimes\mathcal{H}_{2}$ admits the Schmidt
decomposition%
\begin{equation}
|\psi\rangle=\sum_{1\leq k\leq r_{sch}^{(\psi)}}\sqrt{\lambda_{k}(\psi)\text{
}}|e_{k}^{(1)}\rangle\otimes|e_{k}^{(2)}\rangle,\text{ \ \ }\sum_{1\leq k\leq
r_{sch}^{(\psi)}}\lambda_{k}(\psi)=1,\label{22}%
\end{equation}
where each eigenvalue of the reduced states is taken in this sum so many times
what is its multiplicity and $|e_{k}^{(n)}\rangle\in\mathcal{H}_{n}$, $n=1,2,$
are the normalized eigenvectors of the reduced states on $\mathcal{H}_{n}$ of
a pure state $|\psi\rangle\langle\psi|$ on $\mathcal{H}_{1}\otimes
\mathcal{H}_{2}$.

Parameters $\sqrt{\lambda_{k}(\psi)\text{ }}$and $1\leq r_{sch}^{(\psi)}%
\leq\min\{d_{1},d_{2}\}$ are called the Schmidt coefficients and the Schmidt
rank of $|\psi\rangle,$ respectively. For a separable pure bipartite state,
the Schmidt rank equals to $1$.

From Eqs. (27)--(30) in \cite{e.r.loubenets5} it follows and it is also easy
to check that the self-adjoint trace class operators
\begin{align}
T_{1\times S_{2}}^{(\psi)}  &  :=\sum_{k,k_{1}}\sqrt{\lambda_{k}(\psi
)\lambda_{k_{1}}(\psi)}|e_{k}^{(1)}\rangle\langle e_{k_{1}}^{(1)}|\otimes
W_{kk_{1}}^{(2,S_{2})},\label{24}\\
T_{S_{1}\times1}^{(\psi)}  &  :=\sum_{k,k_{1}}\sqrt{\lambda_{k}(\psi
)\lambda_{k_{1}}(\psi)}W_{kk_{1}}^{(1_{,}S_{1})}\otimes|e_{k}^{(2)}%
\rangle\langle e_{k_{1}}^{(2)}|,\nonumber
\end{align}
on $\mathcal{H}_{1}\otimes\mathcal{H}_{2}^{\otimes S_{2}}$ and $\mathcal{H}%
_{1}^{\otimes S_{1}}\otimes\mathcal{H}_{2},$ respectively, where
\begin{align}
W_{kk}^{(n,Sn)}  &  :=\left(  |e_{k}^{(n)}\rangle\langle e_{k}^{(n)}|\right)
^{\otimes S_{n}},\label{25}\\
W_{k\neq k_{1}}^{(n,S_{n})}  &  :=\frac{\left(  |e_{k}^{(n)}+e_{k_{1}}%
^{(n)}\rangle\langle e_{k}^{(n)}+e_{k_{1}}^{(n)}|\right)  ^{\otimes S_{n}}%
}{2^{S_{n}+1}}-\frac{\left(  |e_{k}^{(n)}-e_{k_{1}}^{(n)}\rangle\langle
e_{k}^{(n)}-e_{k_{1}}^{(n)}|\right)  ^{\otimes S_{n}}}{2^{S_{n}+1}}\nonumber\\
&  +i\frac{\left(  |e_{k}^{(n)}+ie_{k_{1}}^{(n)}\rangle\langle e_{k}%
^{(n)}+ie_{k_{1}}^{(n)}|\right)  ^{\otimes S_{N}}}{2^{S_{n}+1}}-i\frac{\left(
|e_{k}^{(n)}-ie_{k_{1}}^{(n)}\rangle\langle e_{k}^{(n)}-ie_{k_{1}}%
^{(n)}|\right)  ^{\otimes S_{n}}}{2^{Sn+1}},\nonumber\\
n  &  =1,2,\nonumber
\end{align}
constitute, correspondingly, the $1\times S_{2}$-setting and $S_{1}\times
1$-setting source operators of a pure bipartite state $|\psi\rangle\langle
\psi|$.

Specified for a bipartite case, Eq. (31) in \cite{e.r.loubenets5} implies that
for either of source operators in (\ref{24}), the trace norms
\begin{equation}
\left\Vert T_{1\times S_{2}}^{(\psi)}\right\Vert _{1},\left\Vert
T_{S_{1}\times1}^{(\psi)}\right\Vert _{1}\leq2\left(  \sum_{k}\sqrt
{\lambda_{k}(\psi)}\right)  ^{2}-1, \label{26}%
\end{equation}
where the right hand sides do not depend on numbers $S_{1},S_{2}$ of
measurement settings at each of two sites. Note that
\begin{equation}
2\left(  \sum_{k}\sqrt{\lambda_{k}(\psi)}\right)  ^{2}-1\leq2r_{sch}^{(\psi
)}-1. \label{27}%
\end{equation}

In view of relations (\ref{15}),(\ref{26}) and (\ref{27}) , bound (\ref{0_4}),
we derive the following new result.

\begin{theorem}
For an arbitrary pure bipartite state $|\psi\rangle\in\mathcal{H}_{1}%
\otimes\mathcal{H}_{2},$ the maximal violation $\mathrm{\Upsilon}_{S_{1}\times
S_{2}}^{(|\psi\rangle\langle\psi|)}$ of $S_{1}\times S_{2}$-setting general
Bell inequalities for any number/type of outcomes at each site admits the
upper bound%
\begin{align}
\mathrm{\Upsilon}_{S_{1}\times S_{2}}^{(|\psi\rangle\langle\psi|)}  &
\leq2\min\left\{  \left(  \sum_{k}\sqrt{\lambda_{k}(\psi)}\right)  ^{2},\text{
}S_{1},S_{2}\right\}  -1\label{28}\\
&  \leq2\min\left\{  r_{sch}^{(\psi)},\text{ }S_{1},S_{2}\right\}  -1,
\label{29}%
\end{align}
where $\lambda_{k}(\psi)$ and $r_{sch}^{(\psi)}$ are, correspondingly, the
Schmidt coefficients and the Schmidt rank of a pure bipartite state
$|\psi\rangle.$
\end{theorem}

In view of the relation%
\begin{align}
&  2\min\left\{  r_{sch}^{(\psi)},\text{ }S_{1},S_{2}\right\}  -1\label{30}\\
&  \leq2\min\{d_{1},d_{2},S_{1},S_{2}\}-1,\nonumber
\end{align}
the upper bounds (\ref{28}), (\ref{29}) are consistent with the general upper
bound (\ref{0_1}) valid for all states, pure or mixed.

Theorem 3 and relation (\ref{16_}) imply

\begin{corollary}
For a pure bipartite state $|\psi\rangle\langle\psi|$, possibly infinite
dimensional, with a finite sum of Schmidt coefficients, the maximal violation
$\mathrm{\Upsilon}_{|\psi\rangle\langle\psi|}$ of all general Bell
inequalities admits the bound%
\begin{equation}
\mathrm{\Upsilon}_{|\psi\rangle\langle\psi|}\leq2\left(  \sum_{k}\sqrt
{\lambda_{k}(\psi)}\right)  ^{2}-1. \label{30_}%
\end{equation}

\end{corollary}

Let some vectors $|\psi_{1}\rangle,|\psi_{2}\rangle\in\mathcal{H}$ be
normalized and linear independent. For each of the following pure entangled
states having the Bell states like forms%
\begin{align}
|\Psi_{00}\rangle &  =\frac{|\psi_{1}\rangle\otimes|\psi_{1}\rangle+|\psi
_{2}\rangle\otimes|\psi_{2}\rangle,}{\sqrt{2\left(  1+\left\vert \langle
\psi_{1}|\psi_{2}\rangle\right\vert ^{2}\right)  }},\label{35}\\
|\Psi_{01}\rangle &  =\frac{|\psi_{1}\rangle\otimes|\psi_{2}\rangle+|\psi
_{2}\rangle\otimes|\psi_{1}\rangle,}{\sqrt{2\left(  1+\left\vert \langle
\psi_{1}|\psi_{2}\rangle\right\vert ^{2}\right)  }},\nonumber\\
|\Psi_{10}\rangle &  =\frac{|\psi_{1}\rangle\otimes|\psi_{1}\rangle-|\psi
_{2}\rangle\otimes|\psi_{2}\rangle}{\sqrt{2\left(  1+\left\vert \langle
\psi_{1}|\psi_{2}\rangle\right\vert ^{2}\right)  }},\nonumber\\
|\Psi_{11}\rangle &  =\frac{|\psi_{1}\rangle\otimes|\psi_{2}\rangle-|\psi
_{2}\rangle\otimes|\psi_{1}\rangle}{\sqrt{2\left(  1+\left\vert \langle
\psi_{1}|\psi_{2}\rangle\right\vert ^{2}\right)  }},\nonumber
\end{align}
the Schmidt rank $r_{sch}^{(\Psi_{jk})}=2,$ $j,k=0,1.$ This and Eq. (\ref{29}) imply.

\begin{corollary}
For each of pure entangled states (\ref{35}), possibly infinite dimensional,
the maximal violation (\ref{16_}) of general Bell inequalities satisfies the
relation%
\begin{equation}
\mathrm{\Upsilon}_{|\Psi_{jk}\rangle\langle\Psi_{jk}|}\leq3,\text{
\ \ }j,k=0,1. \label{36}%
\end{equation}

\end{corollary}

This bound is, in particular, true for each of the Bell states $|\beta
_{jk}\rangle$ on $\mathbb{C}^{2}\otimes\mathbb{C}^{2}$.

\section{Example: bipartite coherent states}

In this Section, we proceed to apply the new results of Theorem 3 and
Corollary 1 to finding upper bounds on the maximal violation of general Bell
inequalities by infinite dimensional entangled bipartite coherent states of
the form (\ref{35}):%
\begin{align}
|\Phi_{1}(\alpha)\rangle &  =\frac{|\alpha\rangle\otimes|\alpha\rangle
+|-\alpha\rangle\otimes|-\alpha\rangle}{\sqrt{2(1+{e^{-4\alpha^{2}})}}%
},\label{37}\\
|\Phi_{2}(\alpha)\rangle &  =\frac{|\alpha\rangle\otimes|-\alpha
\rangle+|-\alpha\rangle\otimes|\alpha\rangle}{\sqrt{2(1+{e^{-4\alpha^{2}})}}%
},\nonumber
\end{align}%
\begin{align}
|\Phi_{3}(\alpha)\rangle &  =\frac{|\alpha\rangle\otimes|\alpha\rangle
-|-\alpha\rangle\otimes|-\alpha\rangle}{\sqrt{2(1-{e^{-4\alpha^{2}})}}%
},\label{37_}\\
|\Phi_{4}(\alpha)\rangle &  =\frac{|\alpha\rangle\otimes|-\alpha
\rangle-|-\alpha\rangle\otimes|\alpha\rangle}{\sqrt{2(1-{e^{-4\alpha^{2}})}}%
},\nonumber
\end{align}
where
\begin{equation}
|\pm\alpha\rangle={e^{-\frac{\alpha^{2}}{2}}}\sum_{m=0}^{\infty}\frac
{(\pm\alpha)^{m}}{\sqrt{m!}}|m\rangle\label{38}%
\end{equation}
are the normalized binary coherent states with parameter $\alpha>0$ and
$\{|m\rangle,\ m=0,1,...\}$ are the Fock vectors.

For $\alpha\rightarrow0$, each of bipartite coherent states (\ref{37}) tends
to the product state $|0\rangle\otimes|0\rangle$ whereas states (\ref{37_})
tend to the corresponding Bell states:
\begin{align}
|\Phi_{3}(\alpha)\rangle &  =\frac{\alpha}{\sqrt{\sinh(2\alpha^{2})}}\left\{
|1\rangle\otimes|0\rangle+|0\rangle\otimes|1\rangle+\mathrm{o}(\alpha
)\right\}  \underset{\alpha\rightarrow0}{\rightarrow}|\beta_{01}\rangle
:=\frac{1}{\sqrt{2}}(|1\rangle\otimes|0\rangle+|0\rangle\otimes|1\rangle
),\label{39}\\
|\Phi_{4}(\alpha)\rangle &  =\frac{\alpha}{\sqrt{\sinh(2\alpha^{2})}}\left\{
|1\rangle\otimes|0\rangle-|0\rangle\otimes|1\rangle+\mathrm{o}(\alpha
)\right\}  \underset{\alpha\rightarrow0}{\rightarrow}-|\beta_{11}%
\rangle:=\frac{1}{\sqrt{2}}(|1\rangle\otimes|0\rangle-|0\rangle\otimes
|1\rangle).\nonumber
\end{align}
Here, $\mathrm{o}(\alpha)$ is a high-order term of $\alpha$.

For states (\ref{37}), the \emph{nonzero} eigenvalues of their reduced states
are nondegenerate and read (see in Appendix)
\begin{equation}
\lambda_{\pm}(\Phi_{1}(\alpha))=\lambda_{\pm}(\Phi_{2}(\alpha))=\frac{\left(
1\pm{e^{-2\alpha^{2}}}\right)  }{2(1+{e^{-4\alpha^{2}})}}^{2}\label{40}%
\end{equation}
for all $\alpha>0.$ For each of states (\ref{37_}), the \emph{nonzero}
eigenvalue of the reduced states is equal to $\frac{1}{2}$ for all $\alpha>0$
and has multiplicity $2$. The Schmidt ranks of states (\ref{37}) and
(\ref{37_}) are equal to $2.$ This and Corollary 1 imply.

\begin{proposition}
For each of infinite dimensional bipartite coherent states $|\Phi_{j}%
(\alpha)\rangle,$ $j=1,...4,$ the maximal violation (\ref{30_}) of general
Bell inequalities%
\begin{align}
\mathrm{\Upsilon}_{|\Phi_{j}\rangle\langle\Phi_{j}|}  &  \leq\frac
{3-{e^{-4\alpha^{2}}}}{1+{e^{-4\alpha^{2}}}},\text{ \ \ }j=1,2,
\label{41}\\
\mathrm{\Upsilon}_{|\Phi_{j}\rangle\langle\Phi_{j}|}  &  \leq3,\text{
\ \ \ }j=3,4, \label{42}%
\end{align}
where
\begin{equation}
\frac{3-e^{-4\alpha^2}}{1+e^{-4\alpha^2}}\le 3
\end{equation}
for all $\alpha>0$.

\end{proposition}

The numerical calculation of the analytical upper bound in (\ref{41}) is presented on Fig. 1.

\begin{figure}[ptb]
\center
\includegraphics[width=0.8\textwidth]{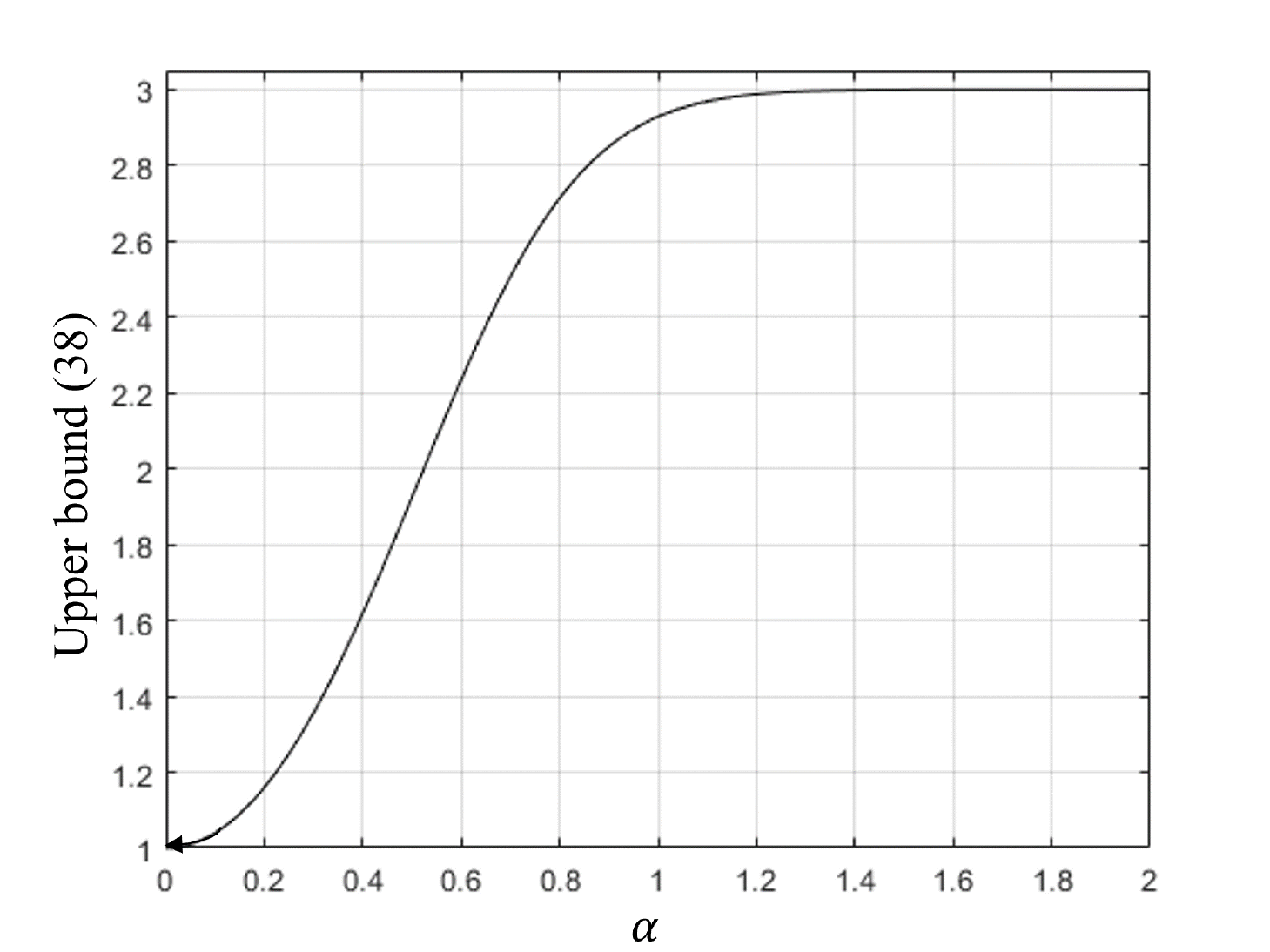}\caption{The
numerical calculation of the analytical upper bound in (\ref{41}).}%
\end{figure}

\section{Conclusion}

In the present article, we find a new upper bound (Theorem 3) on the maximal
violation of general Bell inequalities by a pure bipartite state. This new
upper bound, expressed in terms of the Schmidt coefficients of a pure
bipartite state, is consistent with the upper bound (\ref{0_1}) valid for all
bipartite states, pure or mixed, and implies (Corollary 1) that the maximal
violation of all general Bell inequalities by a pure bipartite state, possibly
infinite dimensional, with a finite sum $\sum_{k}\sqrt{\lambda_{k}(\psi)}$ of
Schmidt coefficients cannot exceed the value $2\left(  \sum_{k}\sqrt
{\lambda_{k}(\psi)}\right)  ^{2}-1.$

As an example, we analyse analytically (Proposition 1) and numerically (see
Fig.1) the upper bounds on the maximal violation of general Bell inequalities
by infinite dimensional bipartite states (\ref{37}), (\ref{37_}), having the
Bell states like forms comprised of two binary coherent states $|\alpha
\rangle,$ $|-\alpha\rangle$ where $\alpha>0.$

\section*{Acknowledgement}

The study by E.R. Loubenets in Sections 2-4 of this work is supported by the
Russian Science Foundation under the Grant 19-11-00086 and performed at the
Steklov Mathematical Institute of Russian Academy of Sciences. The study by
E.R. Loubenets in Section 5 is performed at the National Research University
Higher School of Economics. The study by Min Namkung in Section 5 is performed
at the National Research University Higher School of Economics until August
2021, and at the Kyung Hee University after September 2021. Min Namkung
acknowledges support from the National Research Foundation of Korea (NRF)
grant funded by the Korean government (Ministry of Science and ICT) (NRF2020M3E4A1080088).

\section*{Appendix}

The vectors
\begin{equation}
|u_{1}\rangle:=|\alpha\rangle,\ \ \ |u_{2}\rangle:=\frac{|-\alpha
\rangle-\langle u_{1}|-\alpha\rangle|u_{1}\rangle}{\sqrt{1-|\langle
u_{1}|-\alpha\rangle|^{2}}}\tag{A1}\label{A1}%
\end{equation}
where vector $|u_{2}\rangle$ is due to the Gram-Schmidt orthonormalisation
process between vectors $|\alpha\rangle$ and $|-\alpha\rangle,$ constitute the
orthonormal basis of the linear span of vectors $|\alpha\rangle$ and
$|-\alpha\rangle$. For $\alpha>0$,%
\begin{equation}
|u_{2}\rangle=\frac{|-\alpha\rangle-{e^{-2\alpha^{2}}}|\alpha\rangle}%
{\sqrt{1-{e^{-4\alpha^{2}}}}}\tag{A2}\label{A2}%
\end{equation}
and bipartite coherent states (\ref{37}) and (\ref{37_}) admit the following
decompositions:%
\begin{align}
|\Phi_{1}(\alpha)\rangle &  =\frac{(1+e^{-4\alpha^{2}})|u_{1}\rangle
\otimes|u_{1}\rangle+e^{-2\alpha^{2}}\sqrt{1-{e^{-4\alpha^{2}}}}|u_{1}%
\rangle\otimes|u_{2}\rangle}{\sqrt{2(1+{e^{-4\alpha^{2}})}}}\tag{A3}%
\label{A3}\\
&  +\frac{e^{-2\alpha^{2}}\sqrt{1-{e^{-4\alpha^{2}}}}|u_{2}\rangle
\otimes|u_{1}\rangle+(1-e^{-4\alpha^{2}})|u_{2}\rangle\otimes|u_{2}\rangle
}{\sqrt{2(1+{e^{-4\alpha^{2}})}}},\nonumber\\
|\Phi_{2}(\alpha)\rangle &  =\frac{2e^{-2\alpha^{2}}|u_{1}\rangle\otimes
|u_{1}\rangle+\sqrt{1-e^{-4\alpha^{2}}}\left(  \text{ }|u_{1}\rangle
\otimes|u_{2}\rangle+|u_{2}\rangle\otimes|u_{1}\rangle\right)  }%
{\sqrt{2(1+{e^{-4\alpha^{2}})}}},\nonumber\\
|\Phi_{3}(\alpha)\rangle &  =\frac{(1-e^{-4\alpha^{2}})|u_{1}\rangle
\otimes|u_{1}\rangle-e^{-2\alpha^{2}}\sqrt{1-{e^{-4\alpha^{2}}}}|u_{1}%
\rangle\otimes|u_{2}\rangle}{\sqrt{2(1-{e^{-4\alpha^{2}})}}}\nonumber\\
&  -\frac{e^{-2\alpha^{2}}\sqrt{1-{e^{-4\alpha^{2}}}}|u_{2}\rangle
\otimes|u_{1}\rangle+(1-e^{-4\alpha^{2}})|u_{2}\rangle\otimes|u_{2}\rangle
}{\sqrt{2(1-{e^{-4\alpha^{2}})}}},\nonumber\\
|\Phi_{4}(\alpha)\rangle &  =\frac{1}{\sqrt{2}}(|u_{1}\rangle\otimes
|u_{2}\rangle-|u_{2}\rangle\otimes|u_{1}\rangle).\nonumber
\end{align}
The nonzero eigenvalues of the reduced states of $|\Phi_{j}(\alpha
)\rangle\langle\Phi_{j}(\alpha)|,$ $j=1,2$ can be easily calculated and are
nongenerate and are given by
\begin{equation}
\lambda_{\pm}(\Phi_{j}(\alpha))=\frac{\left(  1\pm{e^{-2\alpha^{2}}}\right)
^{2}}{2(1+{e^{-4\alpha^{2}})}}.\tag{A4}\label{A4}%
\end{equation}
The nonzero eigenvalue of the reduced states of $|\Phi_{j}(\alpha
)\rangle\langle\Phi_{j}(\alpha)|,$  $j=3,4,$ equals to $\frac{1}{2}$ and has
multiplicity $2.$

\end{document}